\documentclass[pdflatex,sn-nature]{sn-jnl}

\usepackage{graphicx}%
\usepackage{multirow}%
\usepackage{amsmath,amssymb,amsfonts}%
\usepackage{amsthm}%
\usepackage{mathrsfs}%
\usepackage[title]{appendix}%
\usepackage{xcolor}%
\usepackage{textcomp}%
\usepackage{manyfoot}%
\usepackage{booktabs}%
\usepackage{algorithm}%
\usepackage{algorithmicx}%
\usepackage{algpseudocode}%
\usepackage{listings}%
\usepackage{subcaption}

\usepackage[inkscapelatex=false]{svg}
\usepackage{makecell}
\usepackage{pdfpages}

\theoremstyle{thmstyleone}%
\theoremstyle{thmstyletwo}%

\theoremstyle{thmstylethree}%

\begin{document}

\title[Article Title]{Efficient E(3)-equivariant framework for universal charge density prediction}


\author[1]{\fnm{Xiwen} \sur{Li}}

\author[1]{\fnm{Zaizhou} \sur{Xin}}

\author[1]{\fnm{Hongyu} \sur{Yu}}

\author[1]{\fnm{Yang} \sur{Zhong}}

\author[1]{\fnm{Xingao} \sur{Gong}}

\author*[1]{\fnm{Hongjun} \sur{Xiang}}\email{hxiang@fudan.edu.cn}

\affil[1]{\orgdiv{Key Laboratory of Computational Physical Sciences (Ministry of Education), 
Institute of Computational Physical Sciences, State Key Laboratory of Surface Physics, 
and Department of Physics}, \orgname{Fudan University}, \orgaddress{\city{Shanghai}, \postcode{200433}, \country{China}}}




\abstract{Electronic structure is ubiquitously obtained via density functional theory (DFT), where the charge density plays a central role. This work presents EdenGNN (Equivariant Density Graph Neural Network), a machine learning (ML) charge density model for electronic structure. Current universal ML charge density models are hampered by prohibitive computational costs. Furthermore, despite being trained on projector augmented-wave (PAW) based DFT datasets, they predict only the pseudo charge density, which is insufficient to reconstruct the electronic structure. In contrast, EdenGNN overcomes these limitations. It additionally predicts the augmentation occupancies, enabling electronic structure calculations with PAW accuracy. Critically, by employing a basis-expansion formulation with fully trainable radial basis functions and a $\Delta$-learning strategy to capture charge transfer, it is over an order of magnitude faster. Trained on the Materials Project database, our universal model, EdenGNN-Uni, accurately predicts the band structures for the majority of materials across a vast chemical space. These findings establish the ML charge density model as a scalable \textit{ab initio} method for large-scale electronic structure calculations and high-throughput screening.}




\maketitle

\section{Introduction}\label{sec1}

Density functional theory (DFT)~\cite{hohenberg_inhomogeneous_1964} is widely used to calculate the electronic structure, which is central to our understanding of materials~\cite{martin2020electronic}. As the fundamental variable in DFT, the charge density is conventionally calculated by solving the Kohn-Sham (KS) equations~\cite{kohn_self-consistent_1965} self-consistently, which is computationally expensive for large-scale calculations. Although early linear-scaling methods~\cite{yang1991direct, wang_charge-density_2002} have successfully demonstrated calculating the electronic structure from approximate charge density in the spirit of the Harris functional~\cite{harris_simplified_1985}, they fail to handle complex chemical environments. With the advent of machine learning (ML) methods in materials science~\cite{behler_generalized_2007,butler2018machine, huang_central_2023, takamoto2022towards, chen2022universal}, the universal ML charge density emerges as a promising approach to address those limitations. A universal ML charge density exhibits not only transferability but also generalizability, avoiding the time-consuming process of preparing DFT datasets and training separately for each class of materials. As a scalable surrogate approach to the KS formalism, the universal ML charge density might accelerate large-scale electronic structure calculations, facilitate new materials discovery, and boost high-throughput materials screening with \textit{ab-initio} accuracy.

However, establishing a universal ML charge density for electronic structure calculations is challenging and remains unsatisfactory. A central difficulty lies in the trade-off between efficiency and generalizability. Learning charge density---a scalar field---requires optimizing values on an enormous number of grid points. While basis-based methods~\cite{brockherde_bypassing_2017, grisafi_transferable_2019, chandrasekaran_solving_2019} focus on optimizing the expansion coefficients instead, their expressivity is restricted by the chosen basis. Existing universal ML charge density models~\cite{koker_higher-order_2024, qin2025eac}, in exchange for generalizability, are grid-based~\cite{gong_predicting_2019, jorgensen_equivariant_2022, focassio_linear_2023} by avoiding the use of an explicit basis set, at the cost of high computational cost for dense real-space grids. Specifically, they directly predict density values using equivariant tensor products, which scale as $\mathcal{O}(L^6)$, where $L$ is the maximum irreducible representation order that is central to the accuracy of the model. A further complication arises when predicting charge density for the projector augmented-wave (PAW)~\cite{blochl_projector_1994} implementation of DFT. The PAW method is broadly employed in materials and chemical sciences because it makes practical the solving of the KS equations with a plane-wave (PW) basis set, achieving a balance between efficiency and accuracy. The charge density representation in the PAW formalism involves not only the pseudo (PS) charge density but also the augmentation occupancies, which constitute the one-center density matrix in the partial wave basis within the augmentation regions. Unlike the scalar PS charge density, augmentation occupancies are tensor quantities transforming covariantly under rotations. Current universal ML charge density models~\cite{koker_higher-order_2024, qin2025eac}, despite being trained on PAW-DFT datasets, omit the augmentation occupancies. As a result, they lack the ingredients to predict the electronic structures in a consistent way. Although this issue has been addressed in~\cite{focassio2024covariant}, the proposed architecture is not suitable for a universal ML model. Notably, while some works have demonstrated universal ML Hamiltonian models~\cite{zhong2024universal, yin2025advancing} for electronic structure, which effectively bypass the need to solve the KS equations, these approaches predict the Hamiltonian matrix elements in the linear combination of atomic orbitals (LCAO) basis, whose accuracy is inherently bounded by the completeness of the chosen basis. 

The contributions of the present work are twofold. First, we introduce EdenGNN (Equivariant Density Graph Neural Network), an ML charge density model formulated for PAW-DFT that predicts the PS charge density and the augmentation occupancies in a unified manner. The novel architecture of EdenGNN employs fully trainable radial basis functions (RBFs) to expand the difference charge density, leveraging the $\Delta$-learning method and thereby achieving both efficiency and generalizability. The transferability of EdenGNN is validated on molecular dynamics trajectories and Moire superlattices. Second, we demonstrate a universal ML charge density called EdenGNN-Uni for electronic structure which is trained on the Materials Project (MP) database~\cite{jain2013commentary}. To the best of our knowledge, EdenGNN-Uni is the first universal ML model for electronic structure predictions with PW accuracy. Through rigorous benchmarks, including extensive band structure comparisons across a diverse materials space, we show that EdenGNN-Uni can capture the electronic structure for the majority of the unseen materials, including both known and novel ones. Our work establishes the ML charge density as a robust approach towards large-scale electronic structure calculations and high-throughput screening, potentially advancing the area of computational materials science.

\section{Results}\label{sec2}

\subsection{Theory}\label{sec:theory}

\begin{figure}[htbp]
    \centering
    \begin{subfigure}[b]{\textwidth}
        \centering
        \includegraphics[width=1.0\linewidth]{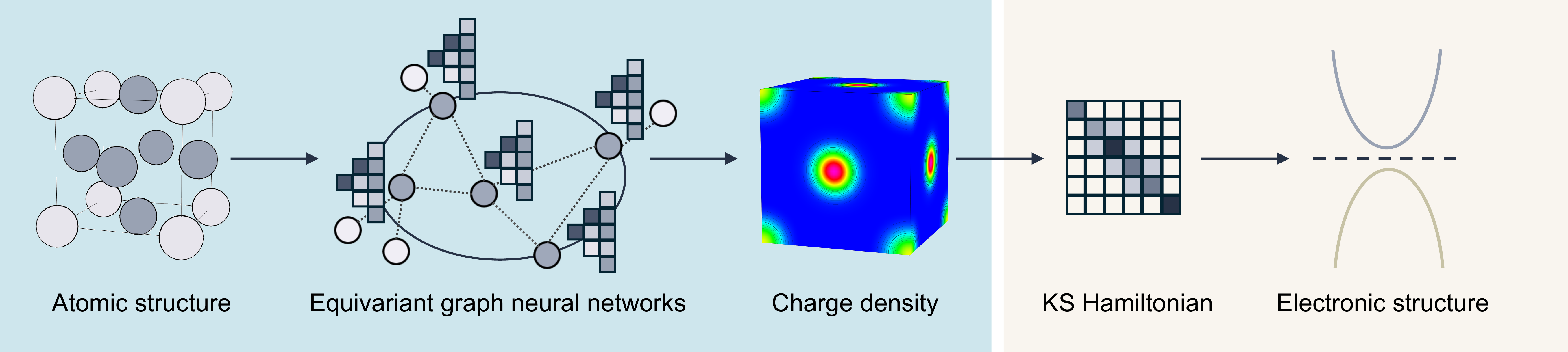}
        \caption{}\label{fig:workflow}
    \end{subfigure}
    \vspace{0.5cm}
    \begin{subfigure}[b]{\textwidth}
        \centering
        \includegraphics[width=\textwidth]{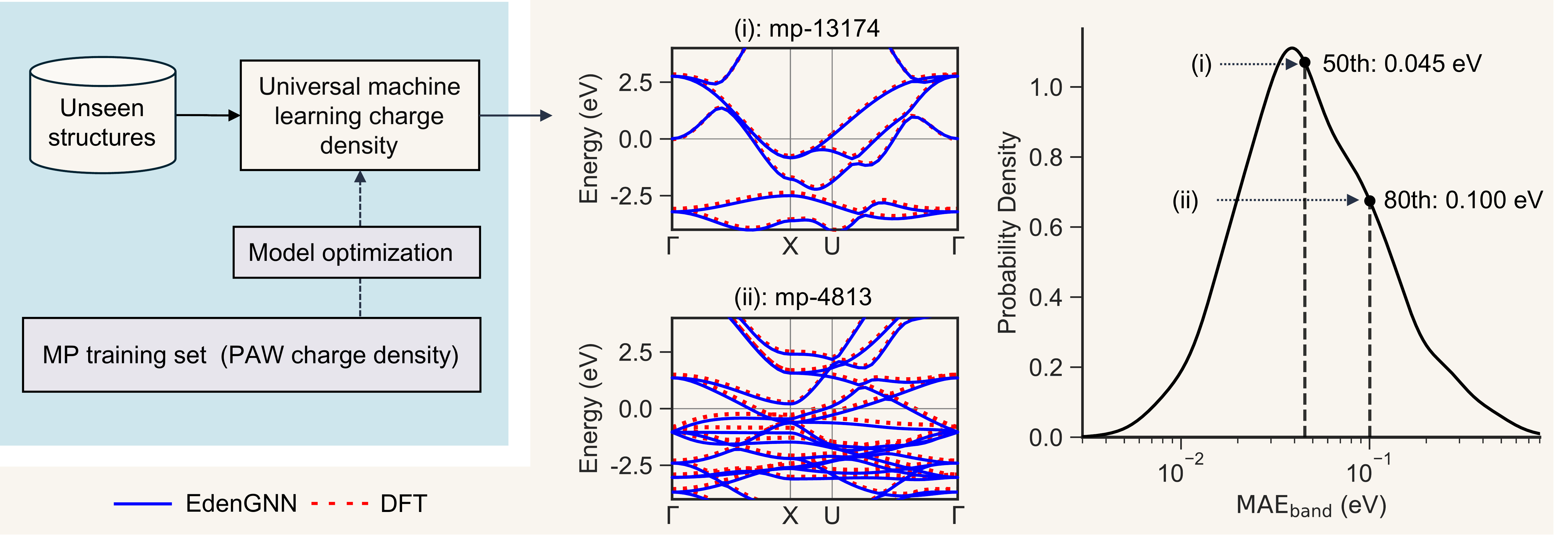}
        \caption{}\label{fig:fidelity}
    \end{subfigure}

    \caption{\textbf{Machine learning charge density for electronic structure calculations.} (a) Schematic of the workflow. Within the framework of graph neural network (GNN), EdenGNN encodes the atomic structure into high-dimensional node features. Exploiting the locality of electronic matter, node features are updated via equivariant message passing algorithm (represented by dashed lines). The final node features output the predicted charge density $n$, from which the Kohn-Sham Hamiltonian $H_{KS}[n]$ is subsequently constructed. Diagonalization of $H_{KS}[n]$ yields the electronic band structure. (b) Evaluation of EdenGNN-Uni. EdenGNN-Uni is trained on structures from the Materials Project database with the projector augmented-wave (PAW) DFT reference data. The right panel illustrates the distribution of the band energy mean absolute error $\mathrm{MAE_{band}}$ in log scale for 1,000 unseen random test structures. $\mathrm{MAE_{band}}$ is evaluated for states above $-4$ eV relative to the Fermi level. Representative band structure comparisons are illustrated for structures in the test set with id (i) mp-13174 and (ii) mp-4813, corresponding to the 50th and 80th percentiles of $\mathrm{MAE_{band}}$ respectively.}
\end{figure}

According to the Hohenberg-Kohn theorem~\cite{hohenberg_inhomogeneous_1964}, the ground-state charge density $n(\mathbf{r})$ uniquely determines the external potential $v(\mathbf{r})$ and therefore the Hamiltonian $H$ and all ground-state state properties of an electronic system. In practice, the central task is to calculate the charge density $n(\mathbf{r})$ given an atomic structure represented by $v(\mathbf{r})$. Traditionally, this is realized through the KS formalism, which incorporates solving the single-particle Schrödinger equation self-consistently for an auxiliary non-interacting system~\cite{kohn_self-consistent_1965}. 

The locality of electronic matter~\cite{kohn_density_1996} provides an alternative possibility of arriving at the charge density. In the absence of long-range electric fields, $n(\mathbf{r})$ only depends on the local chemical environment. This locality enables an $\mathcal{O}(N)$ scaling determination of $n(\mathbf{r})$, which we implement via the statistical learning approach with our EdenGNN architecture---an E(3)-equivariant graph neural network~\cite{thomas_tensor_2018}. As illustrated in Fig.~\ref{fig:workflow}, EdenGNN transforms the atomic structure---a representation of the external potential $v(\mathbf{r})$---into its corresponding charge density. EdenGNN encodes the atomic structure into high-dimensional node features. These features are updated via message-passing algorithm~\cite{gilmer_neural_2017}, which essentially aggregates information from neighbors within a defined cutoff radius to the target nodes. The final charge density is then reconstructed from the chemical environment aware node features. With the approximated charge density $n(\mathbf{r})$, the KS Hamiltonian is constructed in the spirit of the Harris functional~\cite{harris_simplified_1985}. By diagonalizing this Hamiltonian, one can obtain the electronic structure and other related materials properties such as polarization and dielectric function.

EdenGNN is formulated for modeling the charge density of the PAW-PW implementation of DFT. In the PAW formalism~\cite{blochl_projector_1994}, the valence charge density $n(\mathbf{r})$ consists of the PS charge density $\tilde{n}(\mathbf{r})$ and a compensation term inside the augmentation regions represented by the augmentation occupancies $\rho_{AB}$:
\begin{equation}\label{eq:paw_density}
\begin{aligned}
    n(\mathbf{r}) &= \tilde{n}(\mathbf{r}) + \sum_{AB}\rho_{AB}\langle\mathbf{r}|\left(|\phi_B\rangle\langle\phi_A| - |\tilde{\phi}_B\rangle\langle\tilde{\phi}_A|\right)|\mathbf{r}\rangle    \\
    \tilde{n}(\mathbf{r})&=\sum_n \langle\mathbf{r}|\tilde{\Psi}_n\rangle \langle\tilde{\Psi}_n|\mathbf{r}\rangle\\
    \rho_{AB}&=\sum_{n}f_n\langle\tilde{\Psi}_n| \tilde{p}_A\left\rangle\right\langle\tilde{p}_B|\tilde{\Psi}_n\rangle
\end{aligned}
\end{equation}
where $A$ is the composite index denoting the atomic site and the quantum number, and $|\tilde{\Psi}_n\rangle$ are PS wave functions. $|\phi_A\rangle$ are the all-electron wave functions, $|\tilde{\phi}_A\rangle$ are the PS partial waves, and $|p_A\rangle$ are the projector functions. These quantities are fixed once the pseudopotential is generated. Essentially, $\rho_{AB}$ is a density matrix for the one-center expansions in terms of partial waves. $\tilde{n}(\mathbf{r})$ and $\rho_{AB}$ are mixed iteratively during self-consistent calculations which together determine the charge density. 


\subsection{Implementation of EdenGNN: Equivariance, Fully Trainable Radial Basis Functions and Physical Priors}\label{sec:architecture}

\begin{figure}[htbp]
    \centering
    \includegraphics[width=1.0\linewidth]{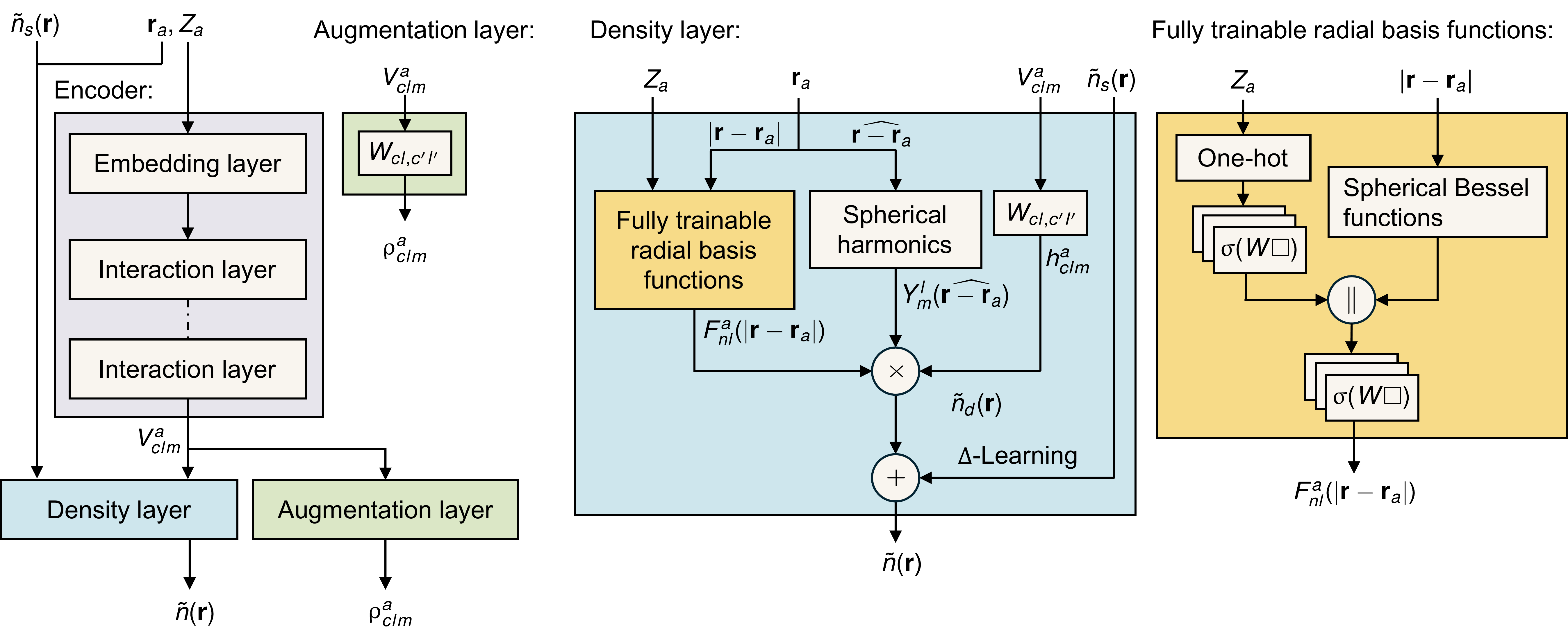}
    \caption{\textbf{EdenGNN model architecture.} $\sigma(W\square)$ denotes multilayer perceptrons (MLPs), $W_{cl,c'l'}$ denotes the equivariant linear transformation and $\parallel$ denotes concatenation. EdenGNN consists of an encoder block, a density layer and an augmentation layer. The encoder transforms the input structure $\mathbf{r}_a, Z_a$ into high-dimensional spherical tensors $V^a_{clm}$. It comprises an embedding layer followed by several interaction layers. The augmentation layer predicts the augmentation occupancies $\rho^a_{clm}$ with an equivariant linear layer. The density layer outputs the pseudo (PS) charge density $\tilde{n}(\mathbf{r})$. A $\Delta$-learning scheme is employed, in which the superposition of atomic charge density $\tilde{n}_s(\mathbf{r})$ serves as a physical prior, and the model directly optimize the difference charge density $\tilde{n}_d(\mathbf{r})$. $\tilde{n}_d(\mathbf{r})$ is expanded with spherical harmonics and radial basis functions as in Eq~.\ref{eq:expansion_trainable}, where the expansion coefficients $h^a_{clm}$ are produced by an equivariant linear layer. The radial basis functions $F^a_{cl}(|\mathbf{r} - \mathbf{r}_a|)$ are fully trainable and parametrized by MLPs, elaborated in the rightmost panel.}\label{fig:architecture}
\end{figure}

Because $\tilde{n}(\mathbf{r})$ and $\rho_{AB}$ are covariant under translation and rotation, in this work we adopt the E(3)-equivariant GNN framework~\cite{thomas_tensor_2018} which has been shown to exhibit transferability in many tasks ranging from machine learning interatomic potentials (MLIPs)~\cite{batzner_e3-equivariant_2022} to ML Hamiltonians~\cite{gong2023general, zhong_transferable_2023}. Reflecting the elegant practice of formulating physical laws in covariant form, an E(3)-equivariant GNN is designed to constrain its predictions to transform equivariantly under E(3) group operations. This is achieved by representing high dimensional hidden features as spherical tensors:
\begin{equation}\label{eq:sph_tensor}
    \sum_{\oplus (l,p)}T^{(l,p)}_m
\end{equation}
where $l$ and $p$ are irreducible representation (IRREP) labels of $O(3)$ group. The IRREP components of spherical tensors are translational invariant and transform under rotation $R$ and inversion $g$ according to their corresponding IRREP of $O(3)$ group: $T^{(l,p)}_m\xrightarrow{R,g} \sigma_p (g)\sum_{m'} D^l_{m'm}(R)T^{l,p}_{m'}$, where $D^l_{m'm}(R)$ are Wigner-D matrices, and $\sigma_p (g) = (-1)^p$. 

In Fig.~\ref{fig:architecture} we illustrate EdenGNN's architecture. EdenGNN consists of an encoder and two output layers. The representation encoder encodes the local chemical environment into high-dimensional node features $V_{clm}^a$ for each atom, and is key to the model's expressiveness and transferability. The representation engineering is well established in the area of MLIPs. Here we adopt the architecture proposed in NequIP, which comprises an embedding layer followed by equivariant interaction layers. Further details can be found in the original work~\cite{batzner_e3-equivariant_2022}. 

The output part consists of a density layer and an augmentation layer, transforming the node features $V_{clm}^a$ into PS charge density and augmentation occupancies, respectively. The augmentation layer follows the standard practice for predicting spherical tensors in E(3)-equivariant GNN. Since they having only onsite terms ($\rho_{AB} =0, \mathbf{r}_a\neq \mathbf{r}_b$), $\rho_{AB}$ are often transformed for practical reasons to spherical tensors $\rho^a_{clm} = \sum_{l_Am_Al_{A'}m_{A'}} C^{lm}_{l_Am_Al_{A'}m_{A'}} \rho_{AA'}$, which directly matches the form of Eq.~(\ref{eq:sph_tensor}). The augmentation layer employs linear operations (denoted by squares) with trainable parameters to couple different IRREP channels equivariantly, as illustrated in Fig.~\ref{fig:architecture}.

The first feature of the density layer is the incorporation of the superposition of atomic charge density $\tilde{n}_s(\mathbf{r})$ as a physical prior. Accordingly the optimization target is the difference charge density:
\begin{equation}\label{eq:difference charge density}
    \tilde{n}_d(\mathbf{r}) = \tilde{n}(\mathbf{r}) - \tilde{n}_s(\mathbf{r})
\end{equation}
Optimizing the charge transfer embodies an application of $\Delta$-learning~\cite{ramakrishnan_big_2015, raissi2019physics}. This is physically motivated because the superposition of atomic charge densities $\tilde{n}_s(\mathbf{r})$ already provides a robust baseline, capturing the majority of the charge distribution. The model can therefore focus its learning capacity on the more subtle and chemically significant charge redistribution that occurs upon bond formation, leading to faster convergence of training and better generalization performance.

Furthermore, for predicting the difference charge density $\tilde{n}_d(\mathbf{r})$, we propose a basis expansion formulation which combines the efficiency of basis-based methods~\cite{grisafi_transferable_2019} and the superior expressiveness of grid-based methods~\cite{koker_higher-order_2024} to reconcile the efficiency-generalizability trade-off:
\begin{equation}\label{eq:expansion_trainable}
    \tilde{n}_d(\mathbf{r}) = \sum_{a} \sum_{clm} h_{clm}^aF^a_{cl}(|\mathbf{r} - \mathbf{r}_a|)Y^l_m(\widehat{\mathbf{r} - \mathbf{r}_a})
\end{equation}
where $h_{clm}^a$ are the expansion coefficients given by a linear transformation of $V_{clm}^a$. The density layer implements Eq.~(\ref{eq:expansion_trainable}), where the RBFs $F^a_{cl}(r)$ are parametrized by multi-layer perceptrons (MLPs) with non-linear activations, capturing a complex, optimal basis across the entire dataset. The essence is that the RBFs $F^a_{cl}(r)$ are fully trainable as the radial filters in the message passing algorithms~\cite{gilmer_neural_2017}. The simple basis expansion formulation of Eq.~(\ref{eq:expansion_trainable}) in fact has connection with the grid-based methods. In the Supplementary Notes, we prove that the two-layer atom-grid equivariant message-passing method which scales as $\mathcal{O}(L^6)$ can be reformulated to an $\mathcal{O}(L^2)$ scaling form, where $L$ is the maximum IRREP order in the equivariant tensor product. It turns out that the nontrivial parts in the equivariant message-passing for scalars are the non-linear gates and the RBFs parameterized by MLPs. Eq.~(\ref{eq:expansion_trainable}) can be seen as a special case where non-linear gates are absent.

\subsection{Validation of EdenGNN's transferability}\label{sec:validation}

Before demonstrating the performance of EdenGNN, we first clarify the error metrics used throughout the work. For the quantities optimized directly by EdenGNN, the PS charge density error is measured by the relative mean absolute error (MAE), normalized by the total number of electrons:
\begin{equation}\label{eq:density_error}
\begin{aligned}
\varepsilon_{\tilde{n}} &= \frac{\int|\hat{\tilde{n}}(\mathbf{r}) - \tilde{n}(\mathbf{r})|d\mathbf{r}}{\int \tilde{n}(\mathbf{r})d\mathbf{r}}  \\
\end{aligned}
\end{equation}
while the augmentation occupancies error is quantified by the MAE, denoted as $\mathrm{MAE_{aug}}$. To evaluate the performance of EdenGNN for electronic structure calculations, the total energy is also compared with the DFT ground-truth~\cite{brockherde_bypassing_2017}. A pragmatic rationale of this requirement is that the total energy serves as the convergence criterion of DFT calculations in practice. In this work, the total energy is accessed by the non-self-consistent calculations of the KS-DFT, and the energy error $\Delta E$ is measured with the MAE metric, denoted by $\mathrm{MAE_{E}}$. We further assess the resulting KS eigenenergies with the MAE metric, denoted by $\mathrm{MAE_{eigen}}$, to illustrate the ML charge density's overall capability in capturing the electronic structure. For evaluating the band structures, we calculate the MAE for states that are above -4 eV relative to the Fermi level, denoted as $\text{MAE}_{\text{band}}$, which directly reflects the level of agreement between the predicted band structures and the DFT benchmarks. This metric is justifiable as the deep core states do not contribute to the low-energy excitations that govern the material's optical and transport properties.


\begin{figure}[htbp]
    \centering
    \begin{subfigure}[b]{0.48\linewidth}
        \centering
        \includegraphics[width=\linewidth]{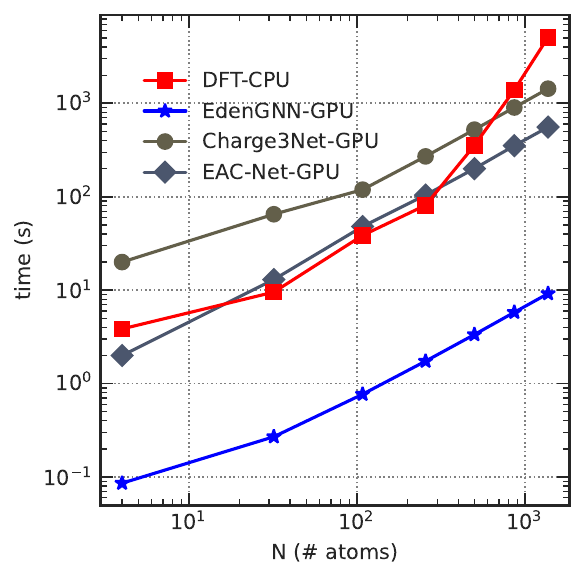}
        \caption{}
        \label{fig:speed}
    \end{subfigure} 
    \hfill
    \begin{subfigure}[b]{0.48\linewidth}
        \centering
        \includegraphics[width=\linewidth]{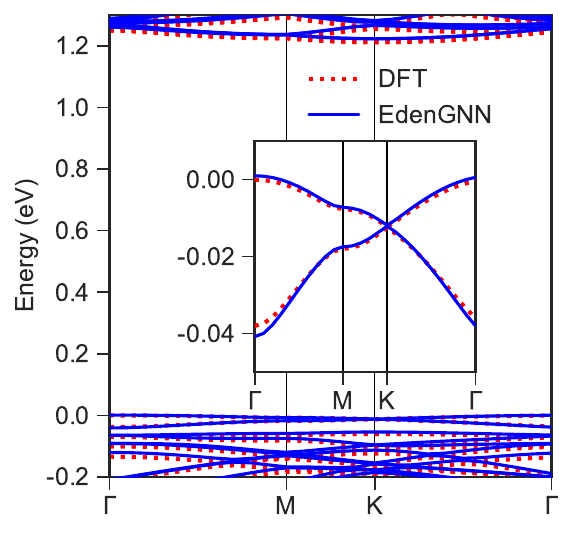}
        \caption{}
        \label{fig:tbm_band}
    \end{subfigure}
    \caption{\textbf{Performance of EdenGNN.} (a) Computational time of obtaining the charge density using DFT and machine learning models, evaluated using various $\mathrm{GaN}$ supercells. (b) Band structures predicted by EdenGNN compared with DFT for a $3.5^{\circ}$ Moiré twisted bilayer $\mathrm{MoS_2}$. Inset: flat-band region. } 
    \label{fig:tbm}
\end{figure}

\subsubsection{Transferability on Molecular Dynamics Trajectories}\label{sec:md}
In this section we demonstrate the transferability of EdenGNN on molecular dynamics (MD) trajectories. We also select two grid-based E(3)-equivariant ML charge density models---Charge3Net~\cite{koker_higher-order_2024} and EAC-Net~\cite{qin2025eac}---to benchmark against EdenGNN. This selection ensures a fair comparison as both claim high accuracy in predicting the PS charge density across diverse applications. For building the ChargE3Net and EAC-Net models we adopt the default model hyperparameter settings from the original works. For training these models we use the same learning rate schedule (see Methods). We select eight representative systems encompassing various bonding characters and symmetries. These include covalently bonded semiconductors such as diamond structure $\mathrm{Si}$, zincblende structured $\mathrm{GaAs}$, $\mathrm{InSb}$ and $\mathrm{InAs}$, and wurtzite structured $\mathrm{GaN}$. We also select ionic rock-salt $\mathrm{LiF}$, which has a large band gap, trigonal $\mathrm{Al_2O_3}$, which has a complex neighbor environment, and cubic $\mathrm{Al}$, which encompasses metallic bonding. For dataset generation, a conventional cell is used for $\mathrm{Al_2O_3}$, while $2\times2\times2$ supercells are used for other systems. The potential energy surface is modeled using CHGNet~\cite{deng_chgnet_2023}, a pre-trained MLIP. For each system, we perform a 250 picoseconds (ps) MD simulation in the NVT ensemble using a Nosé-Hoover (NH) thermostat~\cite{nose_molecular_1984, hoover1985canonical} at 300K. Frames are sampled at 0.5 ps intervals, resulting in a total of 500 frames. To ensure data independence, 100 frames for training and 20 frames for validation are randomly sampled from the first 200 frames, while the final 100 frames are reserved for testing. The distributions of bond lengths for the eight systems are shown in Supplementary Figure S1. For structures such as $\mathrm{InAs}$ and $\mathrm{Al_2O_3}$ the trajectories sample bond length variations of up to 10\%, indicating the MD trajectories produced by CHGNet explore diverse and complex configurations. 

Applying consistent model hyperparameters and optimization strategies (see Methods), we train and evaluate the eight systems individually. As summarized in Table.~\ref{tab:md_traj}, the minute MAEs across all test systems indicate the excellent transferability of EdenGNN. The density and augmentation occupancies exhibit negligible deviations, a result further supported by the parity plots in Supplementary Figure S2. For instance, the relative density error for $\mathrm{Si}$ is merely $0.08\%$ (equivalent to a deviation of only 0.8 electrons per 1000) with a $R^2$ value of 0.99999. Regarding the pseudo charge density, EdenGNN consistently outperforms ChargE3Net and EAC-Net. Furthermore, the eigenenergy errors $\mathrm{MAE_{eigen}}$ for all test systems fall within 20 meV, indicating the capturing of electronic structure. The total energy errors are less than 1 meV per atom, achieving the precision of DFT. In Supplementary Figure S3 we also plot the predicted band structures of these systems, which are barely distinguishable from the DFT benchmarks. The exceptional accuracy achieved in band energies and total energies showcase the fidelity of EdenGNN's predicted charge density. Such robust transferability across unseen MD frames highlights EdenGNN as a powerful tool for investigating the coupling between ionic degrees of freedom and electronic structures.

\begin{table}[htbp]
\caption{\textbf{Performance of ML models on unseen frames of MD trajectories.}}
\begin{tabular}{lcccccc}
    \toprule
    & \multicolumn{4}{c}{EdenGNN} 
    & \multicolumn{1}{c}{ChargE3Net} 
    & \multicolumn{1}{c}{EAC-Net} \\    
    \cmidrule(lr){2-5} \cmidrule(lr){6-6} \cmidrule(lr){7-7}
    \makecell{Systems \\ ~}
    & \makecell{$\varepsilon_{\tilde{n}}$ \\ (\%)}
    & \makecell{$\mathrm{MAE_{aug}}$ \\ ~}
    & \makecell{$\mathrm{MAE}_E$ \\ (meV per atom)}
    & \makecell{$\mathrm{MAE_{eigen}}$ \\ (meV)} 
    & \makecell{$\varepsilon_{\tilde{n}}$ \\ (\%)}
    & \makecell{$\varepsilon_{\tilde{n}}$ \\ (\%)}    
    \\
    \midrule
    $\mathrm{Si}$ & 0.08 & 0.0007 & 0.003 & 0.5 & 0.09 & 0.12\\
    $\mathrm{GaN}$ & 0.04 & 0.0002 & 0.2  & 17.0 & 0.19 & 0.40 \\
    $\mathrm{GaAs}$ & 0.06 & 0.0003 & 0.014 & 3.8  & 0.14 & 0.29 \\
    $\mathrm{Al_2O_3}$ & 0.06 & 0.0003 & 0.02 & 2.8 & 0.13 & 0.21 \\
    $\mathrm{InSb}$ & 0.16 & 0.0010 & 0.4 & 5.4 & 0.24 & 0.45 \\
    $\mathrm{InAs}$ & 0.08 & 0.0007 & 0.047 & 2.8 & 0.22 & 0.25 \\
    $\mathrm{LiF}$ & 0.03 & 0.0001 & 0.006 & 2.2 & 0.17 & 0.31 \\
    $\mathrm{Al}$ & 0.13 & 0.0004 & 0.01 & 0.7 & 0.16 &  0.27 \\
    \bottomrule
\end{tabular}
\footnotetext{$\varepsilon_{\tilde{n}}$ denotes the normalized pseudo charge density error defined in Eq.~(\ref{eq:density_error}). $\mathrm{MAE_{aug}}$, $\mathrm{MAE}_E$ and $\mathrm{MAE_{eigen}}$ represent the mean absolute errors for augmentation occupancies, total energy and eigenenergies, respectively. The augmentation occupancies are dimensionless. }
\label{tab:md_traj}   
\end{table}

As illustrated in Fig.~\ref{fig:speed} we further compare the computational efficiency of ML charge density models and PW DFT in obtaining the charge density. Since the inference time of PS charge density is the primary computational expense, this comparison is fair although EdenGNN additionally predicts the augmentation occupancies compared with ChargE3Net and EAC-Net. The test systems consist of $\mathrm{GaN}$ supercells ranging from $1\times1\times1$ to $7\times7\times7$. DFT calculations are performed on a 64-core Intel Xeon CPU Max 9462 node, whereas ML predictions are carried out on an NVIDIA A800-SXM4-80GB GPU. For large scale systems containing thousands of atoms, EdenGNN is nearly 3 orders of magnitude faster than PW DFT. Furthermore, EdenGNN is around 150 times faster than ChargE3Net and roughly 50 times faster than EAC-Net, demonstrating the supreme efficiency of using fully trainable RBFs in place of equivariant atom-to-grid messages. Although ChargE3Net and EAC-Net are linear scaling, they exhibit large computational prefactors. For instance, for systems with fewer than 100 atoms---where PW DFT remains sufficiently efficient---ChargE3Net is slower than PW DFT and EAC-Net shows no substantial efficiency advantage. In contrast, EdenGNN has a small prefactor and is at least 40 times faster than PW DFT even for small systems, enabling the acceleration of DFT calculations across all scales. The use of different hardware (namely CPU for DFT and GPU for ML models) reflects the corresponding preferences in practical research. The several orders of magnitude difference in scaling and wall time, especially for large systems, robustly establishes the computational advantage of our method over DFT self-consistent calculations.



\subsubsection{Applications to Moire Superlattice}\label{sec:tbm}

We further validate EdenGNN's transferability to large-scale calculations using twisted bilayer $\mathrm{MoS_2}$ (TBM) with a Moiré angle of $3.5^{\circ}$ as a representative case. When two-dimensional (2D) materials are vertically stacked with a small twist angle, Moiré patterns emerge, giving rise to fascinating phenomena such as unconventional superconductivity and strongly correlated states~\cite{kennes2021moire}. Simulating these Moiré superlattices (MSLs) typically requires large supercells containing thousands of atoms, posing a formidable computational challenge for traditional DFT calculations. Among the various constituent materials of MSLs, $\mathrm{MoS_2}$---a 2D transition metal dichalcogenide---has been extensively studied due to its promising applications in electronics and optoelectronics~\cite{manzeli20172d}. When forming a Moiré pattern, TBM undergoes a complex structural reconstruction, which significantly modulates its flat bands and localization behavior~\cite{naik2018ultraflatbands}. Therefore achieving accurate predictions for TBM serves as a rigorous benchmark for evaluating the transferability of machine learning electronic structure models in large-scale calculations. 

Starting with an ideal $3\times3\times1$ supercell of bilayer $\mathrm{MoS_2}$ which contains 54 atoms, the dataset is constructed by introducing random structural deformations, yielding a total of 500 untwisted structures, of which 400 are used for training and the remaining for cross-validation. Specifically, the two layers are subjected to random lateral displacements of distances up to 2~\r{A} and interlayer spacing variations up to 0.5 \r{A}. EdenGNN achieves a density error of $0.06\%$ and an $\mathrm{MAE_{aug}}$ of $7.5\times10^{-4}$ on the validation set. We then predict the charge density of TBM with Moiré angle of $3.5^{\circ}$ containing 1626 atoms in the supercell. EdenGNN achieves high precision with a PS charge density error $\varepsilon_n$ of 0.1\% and an augmentation occupancy MAE of $8.9\times10^{-4}$. The resulting energy error of 0.1 meV/atom validates the fidelity of the predicted charge density. As shown in Fig.~\ref{fig:tbm_band}, the energy bands obtained by diagonalizing the Hamiltonian from predicted charge density are in excellent agreement with the DFT result. Notably, EdenGNN faithfully recovers the characteristic flat band behavior of TBM. Furthermore, Supplementary Figure S4 displays the predicted valence band maximum wavefunctions, which exhibit clear spatial localization at the $B^{S/Mo}$ and $B^{Mo/S}$ regions following structural relaxation. This localized distribution is consistent with previous reports~\cite{naik2018ultraflatbands}, confirming EdenGNN's capability to capture emergent Moiré physics. Overall, EdenGNN shows strong transferability for large-scale calculations even with complex structural reconstruction.

\subsection{A Universal ML Charge Density for Electronic Structures}\label{sec:universal}

\begin{figure}[htbp]
    \centering
    \begin{subfigure}[t]{0.9\linewidth}
        \centering
        \includegraphics[width=\textwidth]{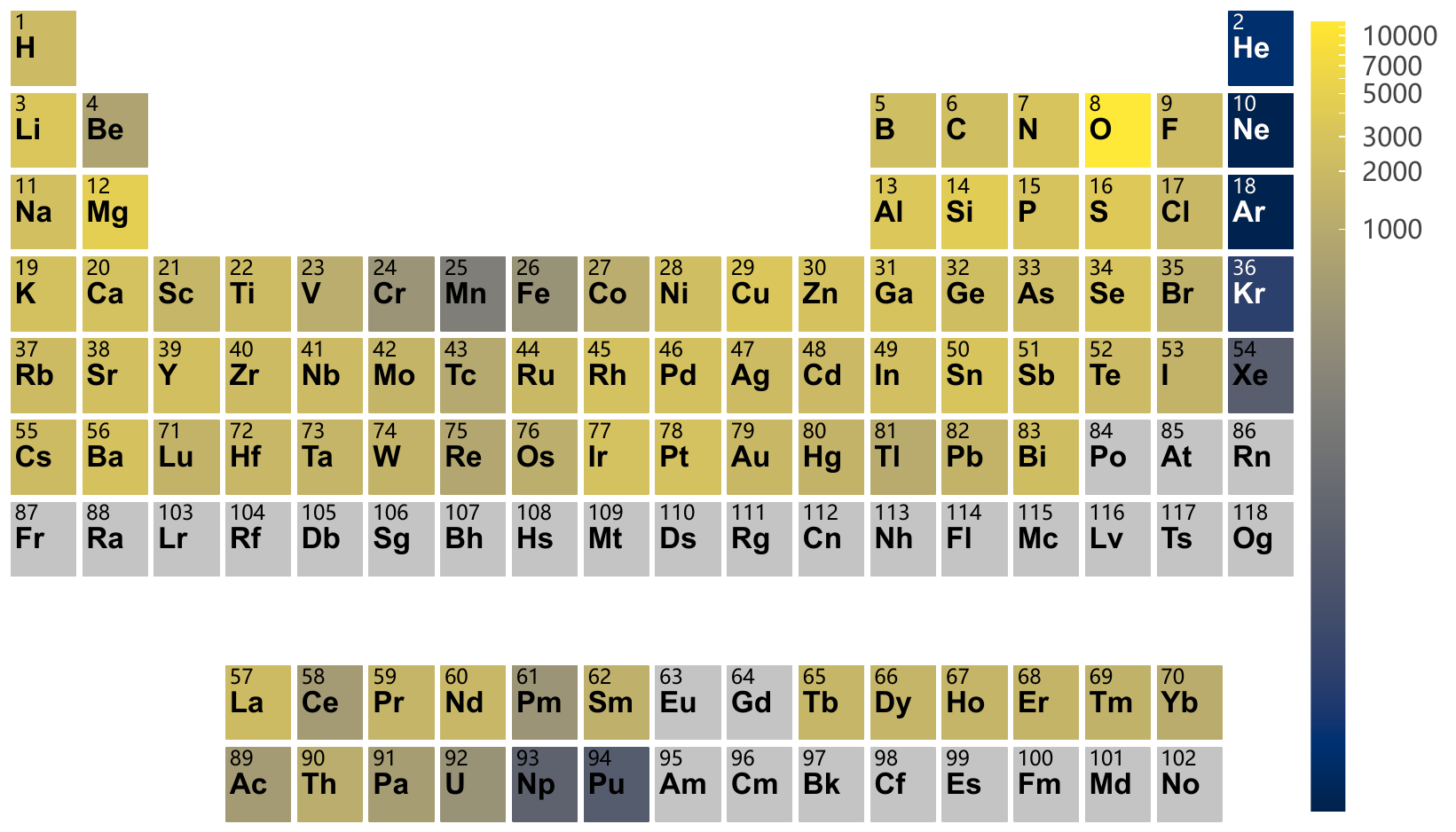}
        \caption{}
        \label{fig:element_count}
    \end{subfigure}
    \vspace{0.5cm}
    \begin{subfigure}[b]{0.3\textwidth}
        \centering
        \includegraphics[width=\textwidth]{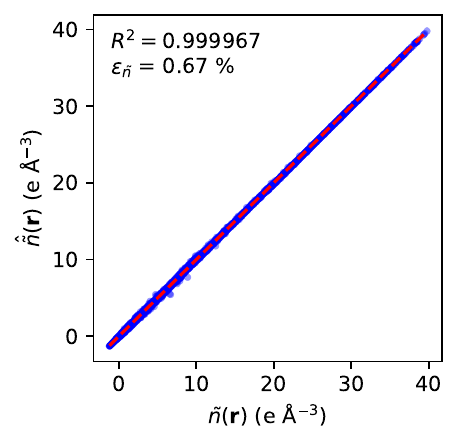}
        \caption{}
        \label{fig:parity_density_mp}
    \end{subfigure}
    \hfill
    \begin{subfigure}[b]{0.3\textwidth}
        \centering
        \includegraphics[width=\textwidth]{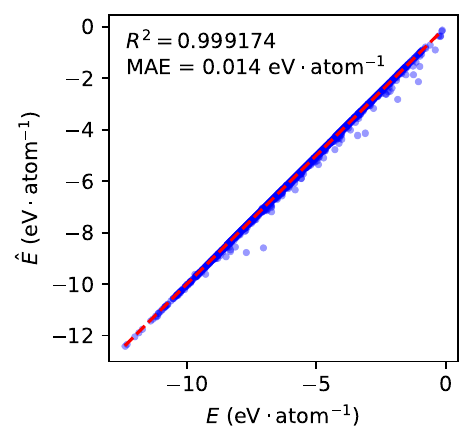}
        \caption{}
        \label{fig:parity_energy_per_atom_mp}
    \end{subfigure}
    \hfill
    \begin{subfigure}[b]{0.3\textwidth}
        \centering
        \includegraphics[width=\textwidth]{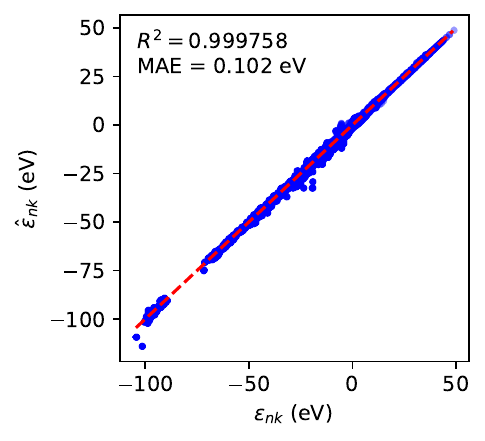}
        \caption{}
        \label{fig:parity_band_energy_mp}
    \end{subfigure}      
    \caption{\textbf{Training of a universal machine learning charge density with EdenGNN.} (a) Element distribution of the training set. All structures are non-magnetic and taken from the Materials Project database. The color bar indicates the total count of atoms per element on a logarithmic scale. (b-d) Parity plots on the test set for:  (b) pseudo (PS) charge density $\tilde{n}(\mathbf{r})$, (c) total energy $E$, and (d) eigenenergy $\varepsilon_{nk}$. The coefficient of determination ($R^2$) and the corresponding error metric are shown in each figure. The error metric $\varepsilon_{\tilde{n}}$ of PS charge density is defined in Eq.~(\ref{eq:density_error}), while other metrics are the mean absolute errors (MAEs).}
    \label{fig:test_universal}
\end{figure}


\begin{figure}[htbp]
    \centering
    \begin{subfigure}[b]{0.45\textwidth}
        \centering
        \includegraphics[width=\textwidth]{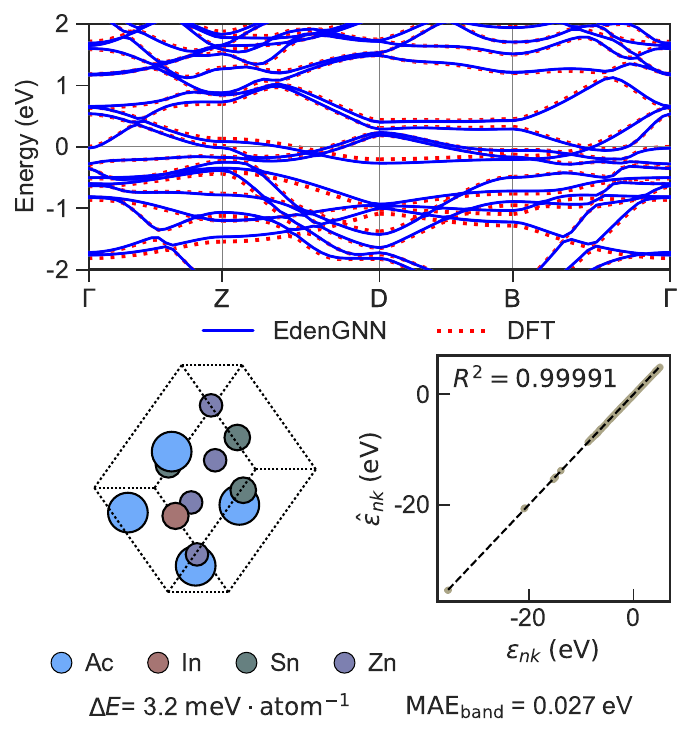}
        \caption{}
    \end{subfigure}
    \hfill
    \begin{subfigure}[b]{0.45\textwidth}
        \centering
        \includegraphics[width=\textwidth]{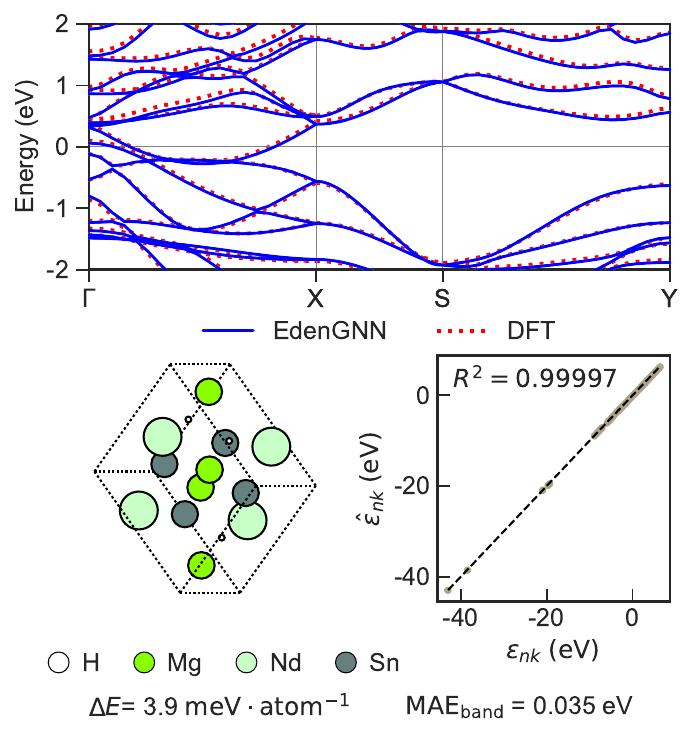}
        \caption{}
    \end{subfigure}
    \vspace{0.5cm}
    \begin{subfigure}[b]{0.45\textwidth}
        \centering
        \includegraphics[width=\textwidth]{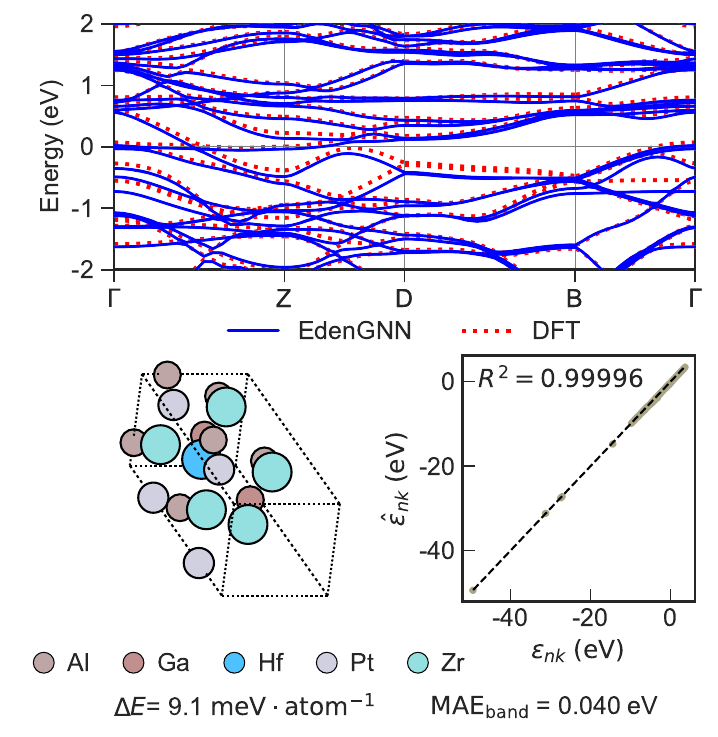}
        \caption{}
    \end{subfigure}
    \hfill
    \begin{subfigure}[b]{0.45\textwidth}
        \centering
        \includegraphics[width=\textwidth]{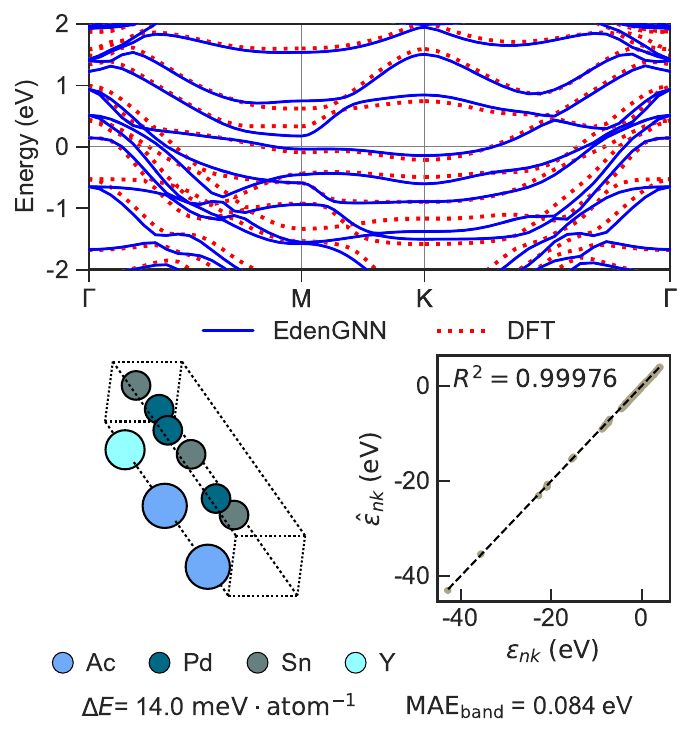}
        \caption{}
    \end{subfigure}    
    \caption{\textbf{Comparison of band structures $\varepsilon_{nk}$ between EdenGNN (red solid lines) and DFT calculations (red dashed lines) for representative test structures from the GNoME database.} For each subfigure, the upper panel illustrates the band structure comparision plot along high-symmetry lines; the lower-left panel displays the atomic struture; and the lower-right panel shows the parity plot of the eigenenergy $\varepsilon_{nk}$ with the coefficient of determination ($R^2$) indicated. The elements of the compound, energy error $\Delta E$ and band energy mean absolute error $\mathrm{MAE_{band}}$ are listed at the bottom. $\mathrm{MAE_{band}}$ is evaluated only for states above -4 eV relative to the Fermi level $E_f$. The representative compounds are: (a) $\mathrm{Ac_4InSn_3Zn_4}$, (b) $\mathrm{H_4Mg_4Nd_4Sn_4}$, (c)$\mathrm{Al_6Ga_2HfPt_4Zr_5}$, and (d)$\mathrm{Ac_2Pd_3Sn_3Y}$.}
    \label{fig:gnome}
\end{figure}

While the success in Moiré systems and MD trajectories confirms EdenGNN's transferability within specific chemical environments, for accelerating high-throughput screening, a universal ML model is required, which must exhibit generalizability across the vast and diverse chemical space. Based on the EdenGNN architecture, in this section we report capturing the electronic structure of the materials space using a universal ML charge density called EdenGNN-Uni. 

The dataset for building EdenGNN-Uni contains 51,000 random non-magnetic materials sampled from the MP database~\cite{jain2013commentary}, spanning 87 elements as depicted in Fig.~\ref{fig:element_count}. The dataset is randomly split into 50,000 and 1,000 for training and cross-validation, respectively. We employ a dual-network architecture: one dedicated to the PS charge density and another to the augmentation occupancies. This is to ensure maximum precision without sacrificing inference efficiency because the overall wall time is governed by the PS density prediction. The training and validation loss curves for EdenGNN-Uni are provided in Supplementary Figure S5.

We first evaluate the performance of EdenGNN-Uni on 5,000 unseen random non-magnetic structures in the MP database, illustrated by the parity plots in Fig.~\ref{fig:test_universal}\subref{fig:parity_density_mp}--\subref{fig:parity_band_energy_mp}. The universal model achieves a normalized MAE of 0.67\% (with the coefficient of determination $\mathrm{R}^2=0.99997$) for the PS charge density and a MAE of 0.0085 ($\mathrm{R}^2=0.99714$) for the augmentation occupancies, demonstrating its remarkable capability to generalize the complex charge transfer across diverse bonding environments. A key advantage of EdenGNN-Uni, compared with previous models such as ChargE3Net and EAC-Net, is that it can predict electronic structures within the PAW formalism. In the following, the predicted total energies and electronic structures are obtained by diagonalizing the Hamiltonian constructed from the predicted charge density. The resulting total energy error of 14 meV/atom ($\mathrm{R}^2=0.99917$) matches the precision of current state-of-the-art universal MLIPs~\cite{riebesell2025framework}, thereby empirically validating the accuracy of EdenGNN-Uni. More importantly, EdenGNN-Uni exhibits exceptional predictive power for electronic structures, as evidenced by the KS eigenenergy MAE of 0.102 eV ($\mathrm{R}^2=0.99976$) in Fig.~\ref{fig:parity_band_energy_mp}. The smaller errors at higher energy levels are expected, as these high-lying scattering states are predominantly governed by the kinetic term and are thus less sensitive to potential variations induced by charge density errors. 

The capability of EdenGNN-Uni for predicting electronic structure is further substantiated through extensive band structure comparisons of 1,000 random sampled test materials, which are available in our open data. Among these, one hundred random comparison plots are plotted in Supplementary Figure S8. Because the errors of universal models typically exhibit a broader distribution than those of  specialized models trained for specific systems, the simple MAEs are insufficient for evaluating their generalizability and fidelity. In Fig.~\ref{fig:fidelity}, we plot the distribution of $\mathrm{MAE_{band}}$ alongside two representative band structures at the 50th and 80th percentiles with MP ID mp-13174 and mp-4813 respectively. As illustrated, the distribution of $\mathrm{MAE_{band}}$ exhibits a standard Gaussian-like curve, indicating that EdenGNN-Uni demonstrates high predictive power for most of the test materials. Crucially, even at the 80th percentile error ($\mathrm{MAE_{band}}=0.100~\mathrm{eV}$), the predicted band dispersions remain highly aligned with the DFT calculations. Significant deviations in the band structure only become apparent for outliers at the tail of the distribution, as exemplified by the 95th percentile (0.232 eV) representative structure shown in Supplementary Figure S6. This is normal and unfortunately inevitable for data-driven methods. Nevertheless, the median $\mathrm{MAE_{band}}$ is a low 0.045 eV, indicating that a typical inferred band structure exhibits quantitative agreement with DFT reference.

To further demonstrate the generalizability of EdenGNN-Uni, we test it on 1,000 novel materials randomly sampled from the GNoME database~\cite{merchant2023scaling}. These compounds exhibit complex element compositions distinct from those in the MP database. As illustrated in the parity plots in Supplementary Figure S7, EdenGNN-Uni achieves a normalized density error $\varepsilon_n$ of 0.66\% ($\mathrm{R}^2=0.99997$), $\mathrm{MAE_{aug}}$ of 0.0105 ($\mathrm{R}^2=0.99863$) and $\mathrm{MAE_{eigen}}$ of 0.099 eV ($\mathrm{R}^2=0.99978$). These errors are consistent with the MP test set results. Representative band structure comparisons are illustrated in Fig.~\ref{fig:gnome}, demonstrating that EdenGNN-Uni accurately captures the electronic structures of those complex GNoME compounds. These results suggest that the ML charge density trained on the MP database successfully generalizes to the GNoME database, indicating its fidelity across diverse data manifolds. Overall, for the vast majority of the test materials across a broad chemical space, the predicted electronic structures agree with DFT, suggesting that EdenGNN-Uni is a robust baseline model.


\section{Discussion}

The cubic scaling complexity of solving the KS equations has long been a central challenge in DFT. To address this issue, we have implemented EdenGNN, an ML charge density model which directly predicts the charge density through statistical inference. In contrast with previous ML charge density models, the novel architecture of EdenGNN overcomes the efficiency-generalizability trade-off in predicting the scalar field of the PS charge density. In addition, EdenGNN predicts the augmentation occupancies required by the PAW formalism, thereby enabling electronic structure calculations with PW accuracy. The demonstrated transferability of EdenGNN highlights its potential for studying systems such as Moiré superlattices and disordered materials. More importantly, trained on the MP database, EdenGNN-Uni captures the electronic structures across a broad materials space, providing a robust data-driven approach for high-throughput screening and new materials discovery. This generalizability in electronic structure prediction is consistent with the early success of linear-scaling methods, in which either the charge density or the density matrix is treated as the central quantity.

Notably, obtaining the electronic structure from predicted charge density unavoidably involves diagonalizing the KS Hamiltonian; however this is not a limitation of ML methods themselves. This difficulty may be alleviated by using the less accurate but much more efficient LCAO basis. Although we validate EdenGNN within PW-basis DFT, its application to LCAO-basis DFT is straightforward. In the LCAO framework, norm-conserving pseudopotentials~\cite{hamann_norm-conserving_1979} are widely adopted~\cite{soler2002siesta}, such that only the pseudo charge density needs to be predicted and the augmentation layer becomes unnecessary. Finally, regarding the ML Hamiltonian approaches~\cite{li_deep-learning_2022}, which directly predict the self-consistent potential to obtain electronic structures, a systematic comparison between these methods and ours is still lacking. Nevertheless, the EdenGNN architecture is general and also provides a direct route toward potential prediction.

\section{Methods}\label{sec13}

\subsection{Datasets Preparation}

All DFT calculations in this work are performed using the Vienna ab initio simulation package (VASP) with projector augmented-wave pseudopotentials~\cite{kresse_efficient_1996, kresse_ultrasoft_1999}. The exchange-correlation functional is treated within the generalized gradient approximation (GGA) parameterized by Perdew, Burke and Ernzerhof (PBE)~\cite{perdew1996generalized}. To reduce GPU memory consumption during training, all datasets were generated using the $\mathrm{PREC = Normal}$ setting.

For the structures in the MD dataset, the plane-wave cutoff energy is set to 520 eV. Gamma-centered k-point sampling is used, with grids of $4\times4\times1$ for $\mathrm{Al_2O_3}$, $3\times3\times2$ for $\mathrm{GaN}$, $3\times3\times3$ for $\mathrm{Al}$ and $2\times2\times2$ for others. For the $\mathrm{MoS_2}$ dataset, the plane-wave cutoff energy is set to the minimum value recommended by the pseudopotentials. We use a $3\times3\times1$ Gamma-centered k-point sampling in the Brillouin zone. The $3.5^{\circ}$ twisted $\mathrm{MoS_2}$ superlattice for validation uses the same calculation setting as in the dataset except for a Gamma-only k-point sampling. For the structures in the MP and GNoME dataset, the plane-wave cutoff energy is set to 520 eV, a Gamma-centered $6\times6\times6$ k-point sampling is used. The parameter $\mathrm{LMAXMIX }$ is set to $6$, specifying the maximum IRREP order of augmentation occupancies to be mixed in self-consistent calculations. 

\subsection{Model Hyperparameters and Training Process} 

EdenGNN is implemented using the PyTorch deep learning framework~\cite{paszke2019pytorch}. Throughout the network a maximum IRREP order of $\mathrm{L_{max}} = 6$ is employed to encode directional information, and the shifted soft plus function $ssp(x) = ln(0.5e^x + 0.5)$ is used as the activation function for the MLPs. The encoder is adapted from NequIP~\cite{batzner_e3-equivariant_2022} with three interaction layers, each consisting of a convolution layer with a residual connection. We use a set of IRREPs of $64\times 0e+64\times 0o+20\times 1o+20\times 1e+16\times 2o+16\times 2e+8\times 3o+8\times 3e+6\times 4o+6\times 4e+5\times 5e+5\times 5o+4\times 6e+4\times 6o$ to embed the node features. Here $5\times 5o$, for example, denotes a direct sum of $5$ tensors of IRREP order $5$ and odd parity. The cutoff of message passing in the encoder is 4 \r{A}.

For the density layer, node features are first equivariantly transformed into expansion coefficients $h^a_{clm}$ via linear mixing between different channels of $V^a_{clm}$, implemented using the e3nn~\cite{geiger_e3nn_2022} library. The number of channels $N_c$ for $h^a_{clm}$ is identical across all IRREPs. We set $N_c = 4$ for the MD and TBM models, and $N_c = 3$ for the universal model. The RBFs are constructed similarly to the radial filters in the encoder. As illustrated in Fig.~\ref{fig:architecture}, a spherical Bessel basis with cosine cutoff function is employed to encode the edge distances~\cite{gasteiger2020directional,unke_physnet_2019}:
\begin{equation}
    \mathbf{H}_n(r) = \sqrt{\frac{2}{r_c}}\frac{\sin(\frac{n\pi r}{r_c})}{r}f_{cutoff}(r)
\end{equation}
where $n$ is the order of the spherical Bessel functions, with $n_{max} = 8$ and $r_c = 4$ \r{A}. Meanwhile, atomic numbers are encoded via one-hot embedding and then fed into an MLP to generate the atomic embeddings $\mathbf{H}_m(Z)$, where $m$ is the embedding dimension (set to $8$). Subsequently, the concatenated atomic and distance embeddings $\mathbf{H}_n(r)\parallel \mathbf{H}_m(Z)$ are fed into an MLP, resulting in the element-dependent radial basis $F^a_{cl}(r)$.

For the augmentation layer, the equivariant linear transformation is similar to that of the density layer. The output representation is the direct sum decomposition  of the direct product of the basis IRREPs $3\times 0e + 2\times 1o + 2\times 2e + 3\times 3o$ with itself, corresponding to the minimal representation required to describe $\rho^a_{clm}$ for all elements using the set of pseudopotentials adopted in this work.

We use the $L_1$ loss for both the densities and the augmentation occupancies. When using a shared encoder to predict $\tilde{n}(\mathbf{r})$ and $\rho^a_{clm}$, as in the case of the MD and TBM models, the total loss is a weighted sum:
\begin{equation}
    \mathcal{L} = w_n ||\hat{\tilde{n}}(\mathbf{r}) - \tilde{n}(\mathbf{r})|| + w_a ||\hat{\rho}^a_{clm} - \rho^a_{clm}||
\end{equation}
where $||\cdot||$ represents the mean absolute error. We employ weights $w_n = 0.9$ and $w_a = 0.1$. All models (including the ChargE3Net and EAC-Net models) are trained using the AdamW optimizer~\cite{loshchilov2017decoupled} with a reduce-on-plateau learning rate schedule. All hyperparameters are detailed in our open-source repository. 


\backmatter

\section*{Code and data availability}

The source code of EdenGNN will be made publicly available once the paper is officially accepted.




\bibliography{sn-bibliography}

\begin{thebibliography}{10}
\expandafter\ifx\csname url\endcsname\relax
  \def\url#1{\burl{#1}}\fi
\expandafter\ifx\csname urlprefix\endcsname\relax\def\urlprefix{URL }\fi
\providecommand{\bibinfo}[2]{#2}
\providecommand{\eprint}[2][]{\url{#2}}
\providecommand{\doi}[1]{\url{https://doi.org/#1}}
\bibcommenthead

\bibitem{hohenberg_inhomogeneous_1964}
\bibinfo{author}{Hohenberg, P.} \& \bibinfo{author}{Kohn, W.}
\newblock \bibinfo{title}{Inhomogeneous {Electron} {Gas}}.
\newblock \emph{\bibinfo{journal}{Physical Review}} \textbf{\bibinfo{volume}{136}}, \bibinfo{pages}{B864--B871} (\bibinfo{year}{1964}).
\newblock \urlprefix\url{https://link.aps.org/doi/10.1103/PhysRev.136.B864}.

\bibitem{martin2020electronic}
\bibinfo{author}{Martin, R.~M.}
\newblock \emph{\bibinfo{title}{Electronic structure: basic theory and practical methods}}  (\bibinfo{publisher}{Cambridge university press}, \bibinfo{year}{2020}).

\bibitem{kohn_self-consistent_1965}
\bibinfo{author}{Kohn, W.} \& \bibinfo{author}{Sham, L.~J.}
\newblock \bibinfo{title}{Self-{Consistent} {Equations} {Including} {Exchange} and {Correlation} {Effects}}.
\newblock \emph{\bibinfo{journal}{Physical Review}} \textbf{\bibinfo{volume}{140}}, \bibinfo{pages}{A1133--A1138} (\bibinfo{year}{1965}).
\newblock \urlprefix\url{https://link.aps.org/doi/10.1103/PhysRev.140.A1133}.

\bibitem{yang1991direct}
\bibinfo{author}{Yang, W.}
\newblock \bibinfo{title}{Direct calculation of electron density in density-functional theory}.
\newblock \emph{\bibinfo{journal}{Physical review letters}} \textbf{\bibinfo{volume}{66}}, \bibinfo{pages}{1438} (\bibinfo{year}{1991}).

\bibitem{wang_charge-density_2002}
\bibinfo{author}{Wang, L.-W.}
\newblock \bibinfo{title}{Charge-{Density} {Patching} {Method} for {Unconventional} {Semiconductor} {Binary} {Systems}}.
\newblock \emph{\bibinfo{journal}{Physical Review Letters}} \textbf{\bibinfo{volume}{88}}, \bibinfo{pages}{256402} (\bibinfo{year}{2002}).
\newblock \urlprefix\url{https://link.aps.org/doi/10.1103/PhysRevLett.88.256402}.

\bibitem{harris_simplified_1985}
\bibinfo{author}{Harris, J.}
\newblock \bibinfo{title}{Simplified method for calculating the energy of weakly interacting fragments}.
\newblock \emph{\bibinfo{journal}{Physical Review B}} \textbf{\bibinfo{volume}{31}}, \bibinfo{pages}{1770--1779} (\bibinfo{year}{1985}).
\newblock \urlprefix\url{https://link.aps.org/doi/10.1103/PhysRevB.31.1770}.

\bibitem{behler_generalized_2007}
\bibinfo{author}{Behler, J.} \& \bibinfo{author}{Parrinello, M.}
\newblock \bibinfo{title}{Generalized {Neural}-{Network} {Representation} of {High}-{Dimensional} {Potential}-{Energy} {Surfaces}}.
\newblock \emph{\bibinfo{journal}{Physical Review Letters}} \textbf{\bibinfo{volume}{98}}, \bibinfo{pages}{146401} (\bibinfo{year}{2007}).
\newblock \urlprefix\url{https://link.aps.org/doi/10.1103/PhysRevLett.98.146401}.

\bibitem{butler2018machine}
\bibinfo{author}{Butler, K.~T.}, \bibinfo{author}{Davies, D.~W.}, \bibinfo{author}{Cartwright, H.}, \bibinfo{author}{Isayev, O.} \& \bibinfo{author}{Walsh, A.}
\newblock \bibinfo{title}{Machine learning for molecular and materials science}.
\newblock \emph{\bibinfo{journal}{Nature}} \textbf{\bibinfo{volume}{559}}, \bibinfo{pages}{547--555} (\bibinfo{year}{2018}).

\bibitem{huang_central_2023}
\bibinfo{author}{Huang, B.}, \bibinfo{author}{Von~Rudorff, G.~F.} \& \bibinfo{author}{Von~Lilienfeld, O.~A.}
\newblock \bibinfo{title}{The central role of density functional theory in the {AI} age}.
\newblock \emph{\bibinfo{journal}{Science}} \textbf{\bibinfo{volume}{381}}, \bibinfo{pages}{170--175} (\bibinfo{year}{2023}).
\newblock \urlprefix\url{https://www.science.org/doi/10.1126/science.abn3445}.

\bibitem{takamoto2022towards}
\bibinfo{author}{Takamoto, S.} \emph{et~al.}
\newblock \bibinfo{title}{Towards universal neural network potential for material discovery applicable to arbitrary combination of 45 elements}.
\newblock \emph{\bibinfo{journal}{Nature Communications}} \textbf{\bibinfo{volume}{13}}, \bibinfo{pages}{2991} (\bibinfo{year}{2022}).

\bibitem{chen2022universal}
\bibinfo{author}{Chen, C.} \& \bibinfo{author}{Ong, S.~P.}
\newblock \bibinfo{title}{A universal graph deep learning interatomic potential for the periodic table}.
\newblock \emph{\bibinfo{journal}{Nature Computational Science}} \textbf{\bibinfo{volume}{2}}, \bibinfo{pages}{718--728} (\bibinfo{year}{2022}).

\bibitem{brockherde_bypassing_2017}
\bibinfo{author}{Brockherde, F.} \emph{et~al.}
\newblock \bibinfo{title}{Bypassing the {Kohn}-{Sham} equations with machine learning}.
\newblock \emph{\bibinfo{journal}{Nature Communications}} \textbf{\bibinfo{volume}{8}}, \bibinfo{pages}{872} (\bibinfo{year}{2017}).
\newblock \urlprefix\url{https://www.nature.com/articles/s41467-017-00839-3}.

\bibitem{grisafi_transferable_2019}
\bibinfo{author}{Grisafi, A.} \emph{et~al.}
\newblock \bibinfo{title}{Transferable {Machine}-{Learning} {Model} of the {Electron} {Density}}.
\newblock \emph{\bibinfo{journal}{ACS Central Science}} \textbf{\bibinfo{volume}{5}}, \bibinfo{pages}{57--64} (\bibinfo{year}{2019}).
\newblock \urlprefix\url{https://pubs.acs.org/doi/10.1021/acscentsci.8b00551}.

\bibitem{chandrasekaran_solving_2019}
\bibinfo{author}{Chandrasekaran, A.} \emph{et~al.}
\newblock \bibinfo{title}{Solving the electronic structure problem with machine learning}.
\newblock \emph{\bibinfo{journal}{npj Computational Materials}} \textbf{\bibinfo{volume}{5}}, \bibinfo{pages}{22} (\bibinfo{year}{2019}).
\newblock \urlprefix\url{https://www.nature.com/articles/s41524-019-0162-7}.

\bibitem{koker_higher-order_2024}
\bibinfo{author}{Koker, T.}, \bibinfo{author}{Quigley, K.}, \bibinfo{author}{Taw, E.}, \bibinfo{author}{Tibbetts, K.} \& \bibinfo{author}{Li, L.}
\newblock \bibinfo{title}{Higher-order equivariant neural networks for charge density prediction in materials}.
\newblock \emph{\bibinfo{journal}{npj Computational Materials}} \textbf{\bibinfo{volume}{10}}, \bibinfo{pages}{161} (\bibinfo{year}{2024}).
\newblock \urlprefix\url{https://www.nature.com/articles/s41524-024-01343-1}.

\bibitem{qin2025eac}
\bibinfo{author}{Qin, X.}, \bibinfo{author}{Lv, T.} \& \bibinfo{author}{Zhong, Z.}
\newblock \bibinfo{title}{Eac-net: Predicting real-space charge density via equivariant atomic contributions}.
\newblock \emph{\bibinfo{journal}{arXiv preprint arXiv:2508.04052}}  (\bibinfo{year}{2025}).

\bibitem{gong_predicting_2019}
\bibinfo{author}{Gong, S.} \emph{et~al.}
\newblock \bibinfo{title}{Predicting charge density distribution of materials using a local-environment-based graph convolutional network}.
\newblock \emph{\bibinfo{journal}{Physical Review B}} \textbf{\bibinfo{volume}{100}}, \bibinfo{pages}{184103} (\bibinfo{year}{2019}).
\newblock \urlprefix\url{https://link.aps.org/doi/10.1103/PhysRevB.100.184103}.

\bibitem{jorgensen_equivariant_2022}
\bibinfo{author}{Jørgensen, P.~B.} \& \bibinfo{author}{Bhowmik, A.}
\newblock \bibinfo{title}{Equivariant graph neural networks for fast electron density estimation of molecules, liquids, and solids}.
\newblock \emph{\bibinfo{journal}{npj Computational Materials}} \textbf{\bibinfo{volume}{8}}, \bibinfo{pages}{183} (\bibinfo{year}{2022}).
\newblock \urlprefix\url{https://www.nature.com/articles/s41524-022-00863-y}.

\bibitem{focassio_linear_2023}
\bibinfo{author}{Focassio, B.}, \bibinfo{author}{Domina, M.}, \bibinfo{author}{Patil, U.}, \bibinfo{author}{Fazzio, A.} \& \bibinfo{author}{Sanvito, S.}
\newblock \bibinfo{title}{Linear {Jacobi}-{Legendre} expansion of the charge density for machine learning-accelerated electronic structure calculations}.
\newblock \emph{\bibinfo{journal}{npj Computational Materials}} \textbf{\bibinfo{volume}{9}}, \bibinfo{pages}{87} (\bibinfo{year}{2023}).
\newblock \urlprefix\url{https://www.nature.com/articles/s41524-023-01053-0}.

\bibitem{blochl_projector_1994}
\bibinfo{author}{Blöchl, P.~E.}
\newblock \bibinfo{title}{Projector augmented-wave method}.
\newblock \emph{\bibinfo{journal}{Physical Review B}} \textbf{\bibinfo{volume}{50}}, \bibinfo{pages}{17953--17979} (\bibinfo{year}{1994}).
\newblock \urlprefix\url{https://link.aps.org/doi/10.1103/PhysRevB.50.17953}.

\bibitem{focassio2024covariant}
\bibinfo{author}{Focassio, B.}, \bibinfo{author}{Domina, M.}, \bibinfo{author}{Patil, U.}, \bibinfo{author}{Fazzio, A.} \& \bibinfo{author}{Sanvito, S.}
\newblock \bibinfo{title}{Covariant jacobi-legendre expansion for total energy calculations within the projector augmented wave formalism}.
\newblock \emph{\bibinfo{journal}{Physical Review B}} \textbf{\bibinfo{volume}{110}}, \bibinfo{pages}{184106} (\bibinfo{year}{2024}).

\bibitem{zhong2024universal}
\bibinfo{author}{Zhong, Y.} \emph{et~al.}
\newblock \bibinfo{title}{Universal machine learning kohn--sham hamiltonian for materials}.
\newblock \emph{\bibinfo{journal}{Chinese Physics Letters}} \textbf{\bibinfo{volume}{41}}, \bibinfo{pages}{077103} (\bibinfo{year}{2024}).

\bibitem{yin2025advancing}
\bibinfo{author}{Yin, S.}, \bibinfo{author}{Dai, Z.}, \bibinfo{author}{Pan, X.} \& \bibinfo{author}{He, L.}
\newblock \bibinfo{title}{Advancing universal deep learning for electronic-structure hamiltonian prediction of materials}.
\newblock \emph{\bibinfo{journal}{arXiv preprint arXiv:2509.19877}}  (\bibinfo{year}{2025}).

\bibitem{jain2013commentary}
\bibinfo{author}{Jain, A.} \emph{et~al.}
\newblock \bibinfo{title}{Commentary: The materials project: A materials genome approach to accelerating materials innovation}.
\newblock \emph{\bibinfo{journal}{APL materials}} \textbf{\bibinfo{volume}{1}} (\bibinfo{year}{2013}).

\bibitem{kohn_density_1996}
\bibinfo{author}{Kohn, W.}
\newblock \bibinfo{title}{Density {Functional} and {Density} {Matrix} {Method} {Scaling} {Linearly} with the {Number} of {Atoms}}.
\newblock \emph{\bibinfo{journal}{Physical Review Letters}} \textbf{\bibinfo{volume}{76}}, \bibinfo{pages}{3168--3171} (\bibinfo{year}{1996}).
\newblock \urlprefix\url{https://link.aps.org/doi/10.1103/PhysRevLett.76.3168}.

\bibitem{thomas_tensor_2018}
\bibinfo{author}{Thomas, N.} \emph{et~al.}
\newblock \bibinfo{title}{Tensor field networks: {Rotation}- and translation-equivariant neural networks for {3D} point clouds} (\bibinfo{year}{2018}).
\newblock \urlprefix\url{http://arxiv.org/abs/1802.08219}.

\bibitem{gilmer_neural_2017}
\bibinfo{author}{Gilmer, J.}, \bibinfo{author}{Schoenholz, S.~S.}, \bibinfo{author}{Riley, P.~F.}, \bibinfo{author}{Vinyals, O.} \& \bibinfo{author}{Dahl, G.~E.}
\newblock \bibinfo{title}{Neural {Message} {Passing} for {Quantum} {Chemistry}} (\bibinfo{year}{2017}).
\newblock \urlprefix\url{http://arxiv.org/abs/1704.01212}.

\bibitem{batzner_e3-equivariant_2022}
\bibinfo{author}{Batzner, S.} \emph{et~al.}
\newblock \bibinfo{title}{E(3)-equivariant graph neural networks for data-efficient and accurate interatomic potentials}.
\newblock \emph{\bibinfo{journal}{Nature Communications}} \textbf{\bibinfo{volume}{13}}, \bibinfo{pages}{2453} (\bibinfo{year}{2022}).
\newblock \urlprefix\url{https://www.nature.com/articles/s41467-022-29939-5}.

\bibitem{gong2023general}
\bibinfo{author}{Gong, X.} \emph{et~al.}
\newblock \bibinfo{title}{General framework for e (3)-equivariant neural network representation of density functional theory hamiltonian}.
\newblock \emph{\bibinfo{journal}{Nature Communications}} \textbf{\bibinfo{volume}{14}}, \bibinfo{pages}{2848} (\bibinfo{year}{2023}).

\bibitem{zhong_transferable_2023}
\bibinfo{author}{Zhong, Y.}, \bibinfo{author}{Yu, H.}, \bibinfo{author}{Su, M.}, \bibinfo{author}{Gong, X.} \& \bibinfo{author}{Xiang, H.}
\newblock \bibinfo{title}{Transferable equivariant graph neural networks for the {Hamiltonians} of molecules and solids}.
\newblock \emph{\bibinfo{journal}{npj Computational Materials}} \textbf{\bibinfo{volume}{9}}, \bibinfo{pages}{182} (\bibinfo{year}{2023}).
\newblock \urlprefix\url{https://www.nature.com/articles/s41524-023-01130-4}.

\bibitem{ramakrishnan_big_2015}
\bibinfo{author}{Ramakrishnan, R.}, \bibinfo{author}{Dral, P.~O.}, \bibinfo{author}{Rupp, M.} \& \bibinfo{author}{Von~Lilienfeld, O.~A.}
\newblock \bibinfo{title}{Big {Data} {Meets} {Quantum} {Chemistry} {Approximations}: {The} {$\Delta$}-{Machine} {Learning} {Approach}}.
\newblock \emph{\bibinfo{journal}{Journal of Chemical Theory and Computation}} \textbf{\bibinfo{volume}{11}}, \bibinfo{pages}{2087--2096} (\bibinfo{year}{2015}).
\newblock \urlprefix\url{https://pubs.acs.org/doi/10.1021/acs.jctc.5b00099}.

\bibitem{raissi2019physics}
\bibinfo{author}{Raissi, M.}, \bibinfo{author}{Perdikaris, P.} \& \bibinfo{author}{Karniadakis, G.~E.}
\newblock \bibinfo{title}{Physics-informed neural networks: A deep learning framework for solving forward and inverse problems involving nonlinear partial differential equations}.
\newblock \emph{\bibinfo{journal}{Journal of Computational physics}} \textbf{\bibinfo{volume}{378}}, \bibinfo{pages}{686--707} (\bibinfo{year}{2019}).

\bibitem{deng_chgnet_2023}
\bibinfo{author}{Deng, B.} \emph{et~al.}
\newblock \bibinfo{title}{{CHGNet} as a pretrained universal neural network potential for charge-informed atomistic modelling}.
\newblock \emph{\bibinfo{journal}{Nature Machine Intelligence}} \textbf{\bibinfo{volume}{5}}, \bibinfo{pages}{1031--1041} (\bibinfo{year}{2023}).
\newblock \urlprefix\url{https://www.nature.com/articles/s42256-023-00716-3}.

\bibitem{nose_molecular_1984}
\bibinfo{author}{Nosé, S.}
\newblock \bibinfo{title}{A molecular dynamics method for simulations in the canonical ensemble}.
\newblock \emph{\bibinfo{journal}{Molecular Physics}} \textbf{\bibinfo{volume}{52}}, \bibinfo{pages}{255--268} (\bibinfo{year}{1984}).
\newblock \urlprefix\url{http://www.tandfonline.com/doi/abs/10.1080/00268978400101201}.

\bibitem{hoover1985canonical}
\bibinfo{author}{Hoover, W.~G.}
\newblock \bibinfo{title}{Canonical dynamics: Equilibrium phase-space distributions}.
\newblock \emph{\bibinfo{journal}{Physical review A}} \textbf{\bibinfo{volume}{31}}, \bibinfo{pages}{1695} (\bibinfo{year}{1985}).

\bibitem{kennes2021moire}
\bibinfo{author}{Kennes, D.~M.} \emph{et~al.}
\newblock \bibinfo{title}{Moir{\'e} heterostructures as a condensed-matter quantum simulator}.
\newblock \emph{\bibinfo{journal}{Nature Physics}} \textbf{\bibinfo{volume}{17}}, \bibinfo{pages}{155--163} (\bibinfo{year}{2021}).

\bibitem{manzeli20172d}
\bibinfo{author}{Manzeli, S.}, \bibinfo{author}{Ovchinnikov, D.}, \bibinfo{author}{Pasquier, D.}, \bibinfo{author}{Yazyev, O.~V.} \& \bibinfo{author}{Kis, A.}
\newblock \bibinfo{title}{2d transition metal dichalcogenides}.
\newblock \emph{\bibinfo{journal}{Nature Reviews Materials}} \textbf{\bibinfo{volume}{2}}, \bibinfo{pages}{1--15} (\bibinfo{year}{2017}).

\bibitem{naik2018ultraflatbands}
\bibinfo{author}{Naik, M.~H.} \& \bibinfo{author}{Jain, M.}
\newblock \bibinfo{title}{Ultraflatbands and shear solitons in moir{\'e} patterns of twisted bilayer transition metal dichalcogenides}.
\newblock \emph{\bibinfo{journal}{Physical review letters}} \textbf{\bibinfo{volume}{121}}, \bibinfo{pages}{266401} (\bibinfo{year}{2018}).

\bibitem{riebesell2025framework}
\bibinfo{author}{Riebesell, J.} \emph{et~al.}
\newblock \bibinfo{title}{A framework to evaluate machine learning crystal stability predictions}.
\newblock \emph{\bibinfo{journal}{Nature Machine Intelligence}} \textbf{\bibinfo{volume}{7}}, \bibinfo{pages}{836--847} (\bibinfo{year}{2025}).

\bibitem{merchant2023scaling}
\bibinfo{author}{Merchant, A.} \emph{et~al.}
\newblock \bibinfo{title}{Scaling deep learning for materials discovery}.
\newblock \emph{\bibinfo{journal}{Nature}}  (\bibinfo{year}{2023}).

\bibitem{hamann_norm-conserving_1979}
\bibinfo{author}{Hamann, D.~R.}, \bibinfo{author}{Schlüter, M.} \& \bibinfo{author}{Chiang, C.}
\newblock \bibinfo{title}{Norm-{Conserving} {Pseudopotentials}}.
\newblock \emph{\bibinfo{journal}{Physical Review Letters}} \textbf{\bibinfo{volume}{43}}, \bibinfo{pages}{1494--1497} (\bibinfo{year}{1979}).
\newblock \urlprefix\url{https://link.aps.org/doi/10.1103/PhysRevLett.43.1494}.

\bibitem{soler2002siesta}
\bibinfo{author}{Soler, J.~M.} \emph{et~al.}
\newblock \bibinfo{title}{The siesta method for ab initio order-n materials simulation}.
\newblock \emph{\bibinfo{journal}{Journal of physics: Condensed matter}} \textbf{\bibinfo{volume}{14}}, \bibinfo{pages}{2745--2779} (\bibinfo{year}{2002}).

\bibitem{li_deep-learning_2022}
\bibinfo{author}{Li, H.} \emph{et~al.}
\newblock \bibinfo{title}{Deep-learning density functional theory {Hamiltonian} for efficient ab initio electronic-structure calculation}.
\newblock \emph{\bibinfo{journal}{Nature Computational Science}} \textbf{\bibinfo{volume}{2}}, \bibinfo{pages}{367--377} (\bibinfo{year}{2022}).
\newblock \urlprefix\url{https://www.nature.com/articles/s43588-022-00265-6}.

\bibitem{kresse_efficient_1996}
\bibinfo{author}{Kresse, G.} \& \bibinfo{author}{Furthmüller, J.}
\newblock \bibinfo{title}{Efficient iterative schemes for \textit{ab initio} total-energy calculations using a plane-wave basis set}.
\newblock \emph{\bibinfo{journal}{Physical Review B}} \textbf{\bibinfo{volume}{54}}, \bibinfo{pages}{11169--11186} (\bibinfo{year}{1996}).
\newblock \urlprefix\url{https://link.aps.org/doi/10.1103/PhysRevB.54.11169}.

\bibitem{kresse_ultrasoft_1999}
\bibinfo{author}{Kresse, G.} \& \bibinfo{author}{Joubert, D.}
\newblock \bibinfo{title}{From ultrasoft pseudopotentials to the projector augmented-wave method}.
\newblock \emph{\bibinfo{journal}{Physical Review B}} \textbf{\bibinfo{volume}{59}}, \bibinfo{pages}{1758--1775} (\bibinfo{year}{1999}).
\newblock \urlprefix\url{https://link.aps.org/doi/10.1103/PhysRevB.59.1758}.

\bibitem{perdew1996generalized}
\bibinfo{author}{Perdew, J.~P.}, \bibinfo{author}{Burke, K.} \& \bibinfo{author}{Ernzerhof, M.}
\newblock \bibinfo{title}{Generalized gradient approximation made simple}.
\newblock \emph{\bibinfo{journal}{Physical review letters}} \textbf{\bibinfo{volume}{77}}, \bibinfo{pages}{3865} (\bibinfo{year}{1996}).

\bibitem{paszke2019pytorch}
\bibinfo{author}{Paszke, A.} \emph{et~al.}
\newblock \bibinfo{title}{Pytorch: An imperative style, high-performance deep learning library}.
\newblock \emph{\bibinfo{journal}{Advances in neural information processing systems}} \textbf{\bibinfo{volume}{32}} (\bibinfo{year}{2019}).

\bibitem{geiger_e3nn_2022}
\bibinfo{author}{Geiger, M.} \& \bibinfo{author}{Smidt, T.}
\newblock \bibinfo{title}{e3nn: {Euclidean} {Neural} {Networks}} (\bibinfo{year}{2022}).
\newblock \urlprefix\url{https://arxiv.org/abs/2207.09453}.

\bibitem{gasteiger2020directional}
\bibinfo{author}{Gasteiger, J.}, \bibinfo{author}{Gro{\ss}, J.} \& \bibinfo{author}{G{\"u}nnemann, S.}
\newblock \bibinfo{title}{Directional message passing for molecular graphs}.
\newblock \emph{\bibinfo{journal}{arXiv preprint arXiv:2003.03123}}  (\bibinfo{year}{2020}).

\bibitem{unke_physnet_2019}
\bibinfo{author}{Unke, O.~T.} \& \bibinfo{author}{Meuwly, M.}
\newblock \bibinfo{title}{{PhysNet}: {A} {Neural} {Network} for {Predicting} {Energies}, {Forces}, {Dipole} {Moments} and {Partial} {Charges}}.
\newblock \emph{\bibinfo{journal}{Journal of Chemical Theory and Computation}} \textbf{\bibinfo{volume}{15}}, \bibinfo{pages}{3678--3693} (\bibinfo{year}{2019}).
\newblock \urlprefix\url{http://arxiv.org/abs/1902.08408}.

\bibitem{loshchilov2017decoupled}
\bibinfo{author}{Loshchilov, I.} \& \bibinfo{author}{Hutter, F.}
\newblock \bibinfo{title}{Decoupled weight decay regularization}.
\newblock \emph{\bibinfo{journal}{arXiv preprint arXiv:1711.05101}}  (\bibinfo{year}{2017}).

\end{thebibliography}

\section*{Acknowledgements}

We acknowledge financial support from the National Key R\&D Program of China (No. 2022YFA1402901), NSFC (grants No. 12188101), Shanghai Science and Technology Program (No. 23JC1400900), the Guangdong Major Project of the Basic and Applied Basic Research (Future functional materials under extreme conditions--2021B0301030005), Shanghai Pilot Program for Basic Research—Fudan University 21TQ1400100 (23TQ017), the robotic AI-Scientist platform of Chinese Academy of Science, and New Cornerstone Science Foundation.

\section*{Author contributions}

H.X. and X.G. supervised the project. X.L. and H.X. proposed the algorithm. X.L. implemented the models, prepared the datasets and performed the analysis with the help of Z.X, H.Y and Y.Z. X.L. and H.X. prepared the manuscript. All authors discussed the results and commented on the manuscript.

\section*{Competing interests}

The authors declare no financial interests.




\includepdf[pages=-]{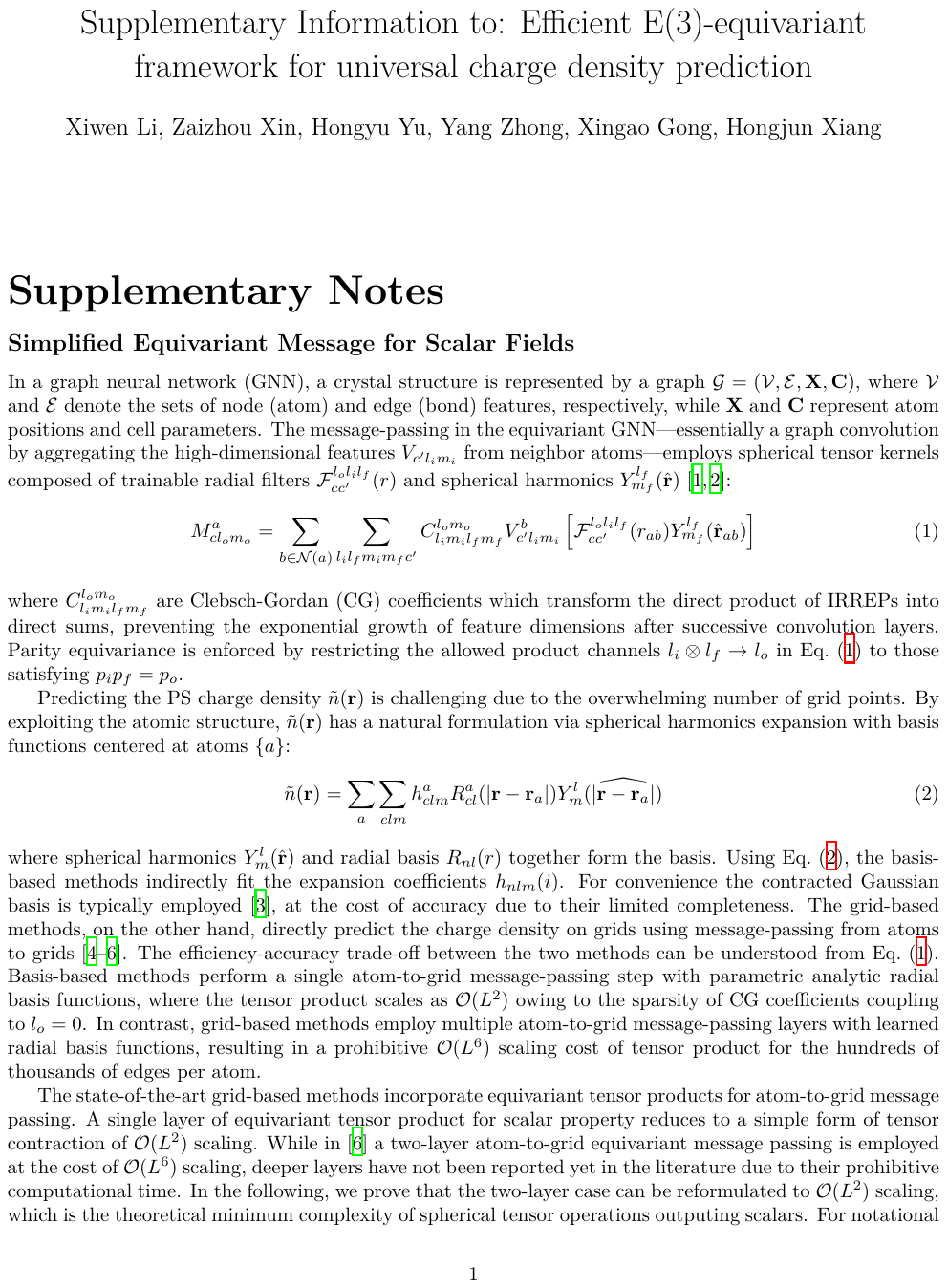}
\end{document}